\documentclass[12pt,a4paper,final]{iopart}

\usepackage{iopams}  
\usepackage{graphicx}
\usepackage[breaklinks=true,colorlinks=true,linkcolor=blue,urlcolor=blue,citecolor=blue]{hyperref}

\expandafter\let\csname equation*\endcsname\relax
\expandafter\let\csname endequation*\endcsname\relax
\usepackage{amsmath}

\usepackage{color}

\begin{document}

\title[Fluctuation relations for fractional Fokker-Planck equations]
      {Fluctuation relations for anomalous dynamics generated by
       time-fractional Fokker-Planck equations}

\author{Peter Dieterich}
\address{Institut f\"ur Physiologie, Medizinische Fakult\"at Carl Gustav Carus, 
         Fetscherstra{\ss}e 74, D-01307 Dresden, Germany}
\ead{peter.dieterich@tu-dresden.de}

\author[cor1]{Rainer Klages}
\address{School of Mathematical Sciences, Queen Mary University of London, 
         Mile End Road, London E1 4NS, UK}
\ead{r.klages@qmul.ac.uk}

\author{Aleksei V. Chechkin$^{1,2,3}$}
\address{$^1$Max-Planck-Institut f\"ur Physik komplexer Systeme, 
             N\"othnitzer Stra{\ss}e 38, D-01187 Dresden, Germany}
\address{$^2$Institute for Theoretical Physics, NSC KIPT, ul.\ Akademicheskaya 1, 
			 UA-61108 Kharkov, Ukraine}
\address{$^3$Institute of Physics and Astronomy, University of Potsdam, D-14476 Potsdam-Golm, Germany}
\ead{achechkin@kipt.kharkov.ua}

\begin{abstract} Anomalous dynamics characterized by non-Gaussian
probability distributions (PDFs) and/or temporal long-range
correlations can cause subtle modifications of conventional
fluctuation relations. As prototypes we study three variants of a
generic time-fractional Fokker-Planck equation with constant
force. Type A generates superdiffusion, type B subdiffusion and
type C both super- and subdiffusion depending on parameter
variation. Furthermore type C obeys a fluctuation-dissipation
relation whereas A and B do not.  We calculate analytically the
position PDFs for all three cases and explore numerically their
strongly non-Gaussian shapes. While for type C we obtain the
conventional transient work fluctuation relation, type A and type B
both yield deviations by featuring a coefficient that depends on
time and by a nonlinear dependence on the work. We discuss possible
applications of these types of dynamics and fluctuation relations to
experiments.  \end{abstract}

\submitto{\NJP}

\section{Introduction}

Understanding fluctuations far from equilibrium defines a key topic of
nonequilibrium statistical physics. A new line of activities started
about three decades ago by discovering different forms of fluctuation
relations (FRs) which generalize fundamental laws of thermodynamics to
small systems in nonequilibrium; see
Refs.~\cite{EvSe02,BLR05,HaSch07,EsHM09,CaHaTa11,JaPiRB11,Seif12,KJJ13}
for reviews and further references therein.  More recently these laws
got unified by over-arching schemes, most notably the deterministic
dissipation function approach by Evans and coworkers
\cite{EvSe02}, and by stochastic thermodynamics
\cite{Maes03,Seki10,Seif12,vdB13}. The latter theory starts from
defining entropy production on the level of individual trajectories in
stochastic models such as Langevin and master equations. Given
that stochastic thermodynamics is based on rather simple Markov models
one may ask to which extent FRs derived from it are reproduced if the
dynamics is more complicated. In our paper we address this problem by
testing FRs for stochastic dynamics that is anomalous due to
non-Markovian dynamical correlations and/or strongly
non-Gaussian PDFs.

Anomalous dynamics has been observed in many experiments and is widely
studied by the theory of anomalous stochastic processes
\cite{MeKl00,CKW04,MeKl04,KRS08,KlSo11,ZDK15}. A characteristic
property of anomalous dynamics is that the mean square displacement
(MSD) grows nonlinearly in time yielding anomalous diffusion in the
long time limit \cite{KRS08}. In contrast, Markovian dynamics like
Brownian motion generates a MSD that increases linearly for long
times. If the MSD grows faster than linear one speaks of
superdiffusion, if it grows slower than linear one obtains
subdiffusion. There are many different ways to model anomalous
stochastic dynamics such as continuous time random walks (CTRW)
\cite{Comp97,CMC97,MKS98,MeKl00}, generalized Langevin equations
\cite{Lutz01,CKW04,Goy12,JCM14}, L\'evy flights and walks
\cite{KBS87,ZDK15}, fractional diffusion equations \cite{KlSo11},
scaled Brownian motion \cite{LiMu02,JeCM14} and heterogeneous
diffusion processes \cite{CCM14}, to name a few.

The study of FRs for anomalous stochastic processes appears to be
rather at the beginning: Crooks and Jarzynski work relations as well
as transient and steady state fluctuation theorems have been confirmed
for non-Markovian Gaussian dynamics modelled by generalized Langevin
equations with memory kernels, given specific conditions are fulfilled
\cite{ZBC05,OhOh07,MaDh07,CCC08}. These results have been reproduced
and generalized by a stochastic thermodynamics approach
\cite{SpSe07}. For non-Gaussian PDFs generated by Langevin equations
with non-Gaussian noise, such as L\'evy noise or Poissonian shot
noise, violations of conventional steady state and transient FRs have
been reported \cite{BeCo04,BaCo09,ToCo07,ToCo09,Budi12,KRG14}. For a
CTRW model with a power law waiting time distribution it was found
that the steady state FRs may or may not hold depending on the
exponent of the waiting time distribution \cite{EsLi08}. Computer
simulations of glassy dynamics exhibiting anomalous diffusion also
showed violations of transient fluctuation relations
\cite{Sell09,CPR13}. In \cite{ChKl09,ChKl12,KCD13} several of the
above types of stochastic dynamics including fractional Fokker-Planck
equations were considered. It was found that the validity of
fluctuation-dissipation relations \cite{KTH92} for a given
anomalous stochastic process plays a crucial role for the validity or
violation of conventional FRs.

In this article we test transient fluctuation relations (TFRs) for a
class of anomalous stochastic processes that so far has not been in
the focus of investigations, which are time-fractional Fokker-Planck
equations (FFPEs). Such equations model the emergence of non-Gaussian
PDFs by using power law memory kernels via time-fractional derivatives
\cite{SKB02}. They need to be distinguished from equations modeling
correlations in space via space-fractional derivatives as they
naturally arise, e.g., for generating L\'evy flights
\cite{MeKl00,ChKl09}.  FFPE can be derived from stochastic equations
of motion either by CTRWs \cite{MeKl00,KlSo11} or by subordinated
Langevin dynamics \cite{Fog94}. Quite a variety of them have been
studied in the literature, both from a purely theoretical point of
view and with respect to applications to experiments: Prominent
examples are fractional Klein-Kramers equations that were used to
analyse biological cell migration data
\cite{BaSi00,MeSo02,DKPS08}. Another type was designed to model the
dynamics of tracer particles in random environments
\cite{FJBE06}. Closely related time-fractional diffusion equations
\cite{Lutz01,SchnWy89,MeKl00} have been used to model a variety of
different processes, from diffusion in crowded cellular environments
\cite{BGM12,KRS08} to geophysical and environmental systems
\cite{MeKl04}. They have also been derived for weakly chaotic
dynamical systems \cite{KCKSG06,KKCSG07}. A bifractional diffusion
equation famously reproduced the spreading of dollar bills in the
United States \cite{BHG06}.

Our paper is structured as follows: In Section~2 we discuss three
types of FFPEs which differ from each other in terms of their
anomalous diffusive properties, and by whether or not they fulfill
fluctuation-dissipation relations. We solve these models for their
position PDFs and study their properties both analytically and
numerically. In Section~3 we test the (work) TFR for our three models
by analytical asymptotic expansions and by numerically plotting the
results. We conclude with a summary and an outlook towards physical
applications in Section~4.

\section{Time-fractional Fokker-Planck equations}

This section introduces to three different types of FFPEs: We first
outline how starting from stochastic dynamics a FFPE generating
superdiffusion can be constructed in the form of an overdamped
Langevin equation with correlated noise. Our argument illustrates how
a time-fractional derivative naturally emerges from modelling power
law time correlation decay. The other two types of FFPEs that we
consider have already been derived in the literature from CTRW theory
and are either subdiffusive or exhibit a transition from sub- to
superdiffusion under parameter variation. We analytically calculate
the first and second moments for all three models, which enables us to
check for the validity of the fluctuation-dissipation relation of the
first kind (FDR1). We also comment on the Galilean invariance of our
models. We then analytically calculate the position PDFs of all FFPEs
and study the solutions numerically by plotting the results.

\subsection{\label{sec:TFDE.Balescu} Constructing a superdiffusive
fractional Fokker-Planck equation}
The study of an overdamped Langevin equation for the position $x(t)$ of a particle on the line driven by 
a correlated stochastic process and an external force allows to gain insight into the origin of a 
superdiffusive FFPE. Our Langevin equation of interest is given by
\begin{equation}
\label{eq:langevin}
\frac{dx}{dt} = \frac{F_0}{m \gamma_\alpha} + v(t),
\end{equation}
where $F_0$ denotes a constant external force, $\gamma_\alpha$ a friction  
coefficient and $m$ the mass of the particle. We assume that $v(t)$ is a stationary correlated stochastic 
process with zero mean $\langle v(t) \rangle = 0$ and a power-law correlation function 
\begin{equation}
\label{eq:vcorrelations}
\langle v(t) \; v(t_1) \rangle = \frac{K_\alpha}{\Gamma(\alpha-1)} \frac{1}{|t-t_1|^{2-\alpha}}
\end{equation}
with $1<\alpha<2$, gamma function $\Gamma$ and generalized diffusion coefficient $K_\alpha$. Note that we do not further specify the noise. Following the pseudo-Liouville hybrid approach of Balescu~\cite{Balescu1997,Bal07} 
(see~\ref{appendix:A}) one obtains the following exact result in Eq.~\ref{eq:f-solution}
for the PDF $f(x,t)$:
\begin{equation}
\left(\frac{\partial}{\partial t} + v_0 \frac{\partial}{\partial x} \right) f(x,t)
= \frac{\partial^2}{\partial x^2} 
\int\limits_0^t dt_1 \; \Big\langle v(t) v(t_1) \;
f\left(x-\Delta(t,t_1),t_1 \right) 
\Big\rangle
\end{equation}
with $v_0 = F_0/(m \gamma_\alpha)$ and
$\Delta(t,t_1)=v_0(t-t_1)+\int\limits_{t_1}^t dt_2 \; v(t_2)$. This
exact equation is non-local in time (i.e.\ non-Markovian) and
non-local in space. We now make a local-in-space approximation by
neglecting the term of the fluctuating displacement $\Delta(t,t_1)$ on
the right hand side of the probability density $f$. Such an
approximation seems to be reasonable in the long time and large space
asymptotic limit if the drift and velocity fluctuations are weak
enough.  This assumption results in the following non-Markovian
Fokker-Planck equation:
\begin{equation}
\label{eq:balescu-non-local-in-space}
\left(\frac{\partial}{\partial t} + v_0 \frac{\partial}{\partial x} \right) f(x,t)
= \frac{\partial^2}{\partial x^2} 
\int\limits_0^t dt_1 \; \langle v(t) v(t_1) \rangle \; f(x,t_1) \: .
\end{equation}
Insertion of the correlation function of velocities Eq.~\ref{eq:vcorrelations} 
into Eq.~\ref{eq:balescu-non-local-in-space} leads to 
\begin{equation}
\label{eq:balescu-non-local-in-space+vcorrelation}
\left(\frac{\partial}{\partial t} + v_0 \frac{\partial}{\partial x} \right) f(x,t)
= \frac{\partial^2}{\partial x^2} 
\frac{K_\alpha}{\Gamma(\alpha-1)} \int\limits_0^t dt_1 \; (t-t_1)^{\alpha-2} \; f(x,t_1)\: .
\end{equation}
%
The integral on the right hand side matches to the Riemann-Liouville (RL) fractional 
integral of order $\mu$ given by \cite{SKM93}
\begin{equation}
\label{eq:RLintegral}
J_t^{\mu} \; g(t) \equiv D_t^{-\mu} g(t) = \frac{1}{\Gamma(\mu)} \int\limits_0^t d\tau \; (t-\tau)^{\mu-1} \; g(\tau) 
\end{equation}
with $\mu > 0$ and $\mu = \alpha-1$ for Eq.~\ref{eq:balescu-non-local-in-space+vcorrelation}. 
We also introduce the definition of the RL fractional derivative of positive order
\begin{equation}
\label{eq:RLderiv}
D_t^{\mu}\; g(t) = \frac{d^n}{dt^n} \; J_t^{n-\mu}\; g(t)
\end{equation}
with $\mu>0\:,\:n = \left[ \mu\right] + 1$, where $\left[ \ldots\right] $ refers to the integer part of the given number.  
Applying Eq.~\ref{eq:RLintegral} to Eq.~\ref{eq:balescu-non-local-in-space+vcorrelation} gives us 
our first type of FFPE that we denote as
\begin{equation}
\label{eq:FFPE-A}
\mbox{type A:}\;\;\;
\frac{\partial f_{A}(x,t)}{\partial t} = 
-\frac{\partial}{\partial x}
\left[
v_0 - K_\alpha D_t^{1-\alpha} \frac{\partial}{\partial x}
\right] f_{A}(x,t)\:,\:1<\alpha<2\:.
\end{equation}
To show the relation of this equation with previous works we put $v_0=0$. Then it can be
written as
\begin{equation}
\label{eq:FDE+RLderiv}
\frac{\partial^2}{\partial t^2} f(x,t) 
= K_\alpha \;\frac{\partial^2}{\partial x^2} \; D_t^{2-\alpha} f(x,t)\:,1<\alpha<2\:.
\end{equation}
This equation was called a fractional wave equation in the seminal
paper of Schneider and Wyss \cite{SchnWy89} and has also been derived
for a long-range correlated dichotomous stochastic process
\cite{WGMN97} from a fractional Klein-Kramers equation \cite{BaSi00}
and from a generalised Chapman-Kolmogorov equation \cite{MeKl00b}.
The solution of this equation has been studied in detail in
\cite{Metzler2000e} where it was called a fractional kinetic equation
for sub-ballistic superdiffusion. The equivalent form of this equation
using the Caputo fractional derivative was investigated in
\cite{GLM00}.

Our presentation above illustrates how a FFPE can be derived from
a Langevin equation with power-law decay in the velocity correlation
function. It furthermore demonstrates that a fractional derivative
provides the natural mathematical formulation to model equations
containing power law memory kernels. 

\subsection{Definition and properties of fractional Fokker-Planck equations}
In addition to type A FFPE Eq.~\ref{eq:FFPE-A} we consider two further
types of FFPEs. Both have been derived from CTRW theory 
\cite{Comp97,CMC97,MKS98,MeKl00}. Note that the underlying stochastic dynamics and the
derivation of these two FFPEs are very different from what we
presented for type A above. Indeed, both type B and type C are essentially
(almost) Markovian models, in contrast to type A. Our two new FFPEs
describe subdiffusion under the influence of a constant external force
and naturally appear in physical systems where diffusion is slowed
down by deep traps \cite{MeKl00,MKS98,GBU15}. The difference
between these two types arises from the position of the fractional RL 
derivative with respect to the diffusive and drift part of the equations and the range
of the anomaly parameter $\alpha$. Our second FFPE is defined as
\begin{equation}
\label{eq:FFPE-B}
\mbox{type B:}\;\;\;
\frac{\partial f_{B}(x,t)}{\partial t} =  
-\frac{\partial}{\partial x}
\left[
v_0 - K_\alpha D_t^{1-\alpha} \frac{\partial}{\partial x}
\right] f_{B}(x,t).
\end{equation}
For type C FFPE the RL fractional derivative is also included in the drift term:
\begin{equation}
\label{eq:FFPE-C}
\mbox{type C:}\;\;\;
\frac{\partial f_{C}(x,t)}{\partial t} =  
-\frac{\partial}{\partial x}
\left[
A_{\alpha}v_0 D_t^{1-\alpha} - K_\alpha D_t^{1-\alpha} \frac{\partial}{\partial x}
\right] f_{C}(x,t) ,
\end{equation}
where $A_{\alpha}$ has a dimension of time to the power of $1-\alpha$.
Note that type B and type C FFPEs are defined for 
$0 < \alpha < 1$ whereas for type A FFPE 
$\alpha$ is in the range $1<\alpha<2$. For all three FFPEs we use the initial condition $f_{A,B,C}(x,t=0)=\delta(x)$. 
By means of Fourier and Laplace transforms
\begin{equation}
\label{eq:FT+LT}
\hat{f}(k) = \int\limits_{-\infty}^{\infty} dx e^{i k x} f(x) ,
\;\;\; \;\;\; 
\tilde{f}(s) = \int\limits_{0}^{\infty} dt e^{- s t} f(t)
\end{equation}
a solution of Eqs.~\ref{eq:FFPE-A}, \ref{eq:FFPE-B} and \ref{eq:FFPE-C} 
can be obtained in Fourier-Laplace space as
\begin{equation}
\label{eq:FFPE-B-FT-LT}
\hat{\tilde{f}}_{A,B}(k,s) = \frac{1}{s + v_0 i k + K_\alpha k^2 s^{1-\alpha}},
\end{equation}
\begin{equation}
\label{eq:FFPE-C-FT-LT}
\hat{\tilde{f}}_{C}(k,s) = \frac{1}{s + A_{\alpha}v_0 i k s^{1-\alpha} + K_\alpha k^2 s^{1-\alpha}},
\end{equation}
where the fractional derivative $D^{1-\alpha}_t f(t)$ transforms to $s^{1-\alpha} \tilde{f}(s)$. 
The solutions of type A and type B
FFPE only differ in the range of $\alpha$ as defined above. The representation
in Fourier-Laplace space allows the calculation of moments by differentiation with respect to
$k$:
\begin{equation}
\langle x^n(t) \rangle = {\cal L}^{-1} \left\lbrace (i)^n
\left. \frac{\partial^n \hat{\tilde{f}}(k,s)}{\partial k^n} \right|_{k = 0} \right\rbrace .
\end{equation}
After Laplace inversion one obtains the first two moments and the central second moment for 
$\delta x = x - \langle x \rangle $ of type C FFPE defined in Eq.~\ref{eq:FFPE-C} \cite{MKS98}
\begin{eqnarray}
\label{eq:FFPE-C-moment1}
\langle x \rangle_{C} & = & \frac{A_{\alpha}v_0 t^\alpha}{\Gamma(\alpha+1)}, \\
\label{eq:FFPE-C-moment2}
\langle x^2 \rangle_{C} & = & \frac{2 K_\alpha t^\alpha}{\Gamma(\alpha+1)} 
                    + \frac{2 A_{\alpha}^2v_0^2 t^{2 \alpha}}{\Gamma(2 \alpha+1)},\\
\label{eq:FFPE-C-moment2c}
\langle (\delta x)^2 \rangle_{C} & = &  \frac{2 K_\alpha t^\alpha}{\Gamma(\alpha+1)} 
+ A_{\alpha}^2v_0^2 t^{2 \alpha} \; \left[ \frac{2}{\Gamma(2 \alpha+1)} - \frac{1}{\Gamma(\alpha+1)^2}\right] .
\end{eqnarray}
These results show that the FDR1 \cite{KTH92,ChKl12}
$\langle x(t)\rangle_C \sim \langle x^2(t)\rangle^{v_0=0}_C$ 
is valid for type C. Interestingly the external force influences the 
second central moment $\sim v_0^2 t^{2\alpha}$. Technically this is due to the coupling
term $v_0 ik s^{1-\alpha}$ in the Laplace-Fourier representation of Eq.~\ref{eq:FFPE-C-FT-LT}.
The first moment increases sublinearly despite the constant external force. This can be
interpreted as a partial sticking effect of particles~\cite{Metzler2000h}. By contrast, the
second central moment shows a crossover from $\sim t^\alpha$ to $\sim t^{2\alpha}$. Thus, 
for $v_0 \neq 0$ type C switches from a subdiffusive behavior of the second central 
moment for $0 < \alpha < 1/2$ to a superdiffusive behavior for $1/2 < \alpha < 1$ \cite{MeKl00}.

Analogously, the moments of type A and type B FFPEs of Eq.~\ref{eq:FFPE-A} and 
Eq.~\ref{eq:FFPE-B} are obtained as \cite{MKS98}
\begin{eqnarray}
\label{eq:FFPE-B-moment1}
\langle x \rangle_{A,B} & = & v_0 t, \\
\label{eq:FFPE-B-moment2}
\langle x^2 \rangle_{A,B} & = & \frac{2 K_\alpha t^\alpha}{\Gamma(\alpha+1)} + v_0^2 t^2, \\
\label{eq:FFPE-B-moment2c}
\langle (\delta x)^2 \rangle_{A,B} & = &  \frac{2 K_\alpha t^\alpha}{\Gamma(\alpha+1)} .
\end{eqnarray}
In both cases the first moment only depends on $v_0$ and increases
linearly in time.  The second central moment shows a superdiffusive
and subdiffusive increase $\sim t^\alpha$ for type A and type B FFPE,
respectively. In contrast to type C FFPE, the second moment of type A
and type B FFPEs is without any coupling to $v_0$.  In addition, type
A and type B FFPEs break FDR1 between the first $\langle
x(t)\rangle_{A,B}$ and the second moment $\langle
x^2(t)\rangle^{v_0=0}_{A,B}$. In both cases this is what one should
expect according to the definition of both models: Type A is based on
the Langevin equation \ref{eq:langevin} where the
fluctuation-dissipation relation of the second kind (FDR2) is broken
by construction. Note that FDR2 establishes a relation between the
noise and the friction \cite{KTH92}.  The breaking of FDR2 suggests a
breaking of FDR1 as was shown for Gaussian stochastic processes in
\cite{ChKl12}.  For type B the fractional derivative acts only on the
diffusion term in Eq.~\ref{eq:FFPE-B} thus breaking FDR1 while for
type C it acts simultaneously on both the drift and the diffusion
terms in Eq.~\ref{eq:FFPE-C} hence preserving FDR1.

A second difference between these FFPEs consists in their behavior
under Galilean transformation. With $X = x - v_0 t$ and $T = t$ the
PDF $f(x,t)$ is transformed to $\Omega(X,T)$. The coupling of the
fractional RL derivative to the $v_0$ drift term of type C FFPE in
Eq.~\ref{eq:FFPE-C} breaks Galilean invariance. However, type A and B
FFPE of Eq.~\ref{eq:FFPE-A} and Eq.~\ref{eq:FFPE-B} fulfill Galilean
invariance in the long time and large space limit
\cite{MKS98,MeKl00,CaBa15}, where they can be written as
\begin{equation}
\label{eq:FDE2nodrift}
\frac{\partial \Omega_{A,B}(X,T)}{\partial T} 
 = K_{\alpha} D^{1-\alpha}_T \frac{\partial^2 \Omega_{A,B}(X,T)}{\partial X^2}\:.\\
\end{equation}
This means that in this limit breaking or preserving FDR1
corresponds to preserving respectively breaking Galilean invariance in
the case of these FFPEs. This property will be exploited in the next
subsection where we discuss analytical and numerical solutions of our
three types of FFPEs.
 
\subsection{Analytical solution of time-fractional Fokker-Planck equations}
\subsubsection*{Type C FFPE:}
Fourier inversion~\cite{ChKl09}
leads to the solution of type C FFPE in $(x,s)$ space:
\begin{equation}
\label{eq:FFPE-C-xs}
\tilde{f}_C(x,s) = \frac{s^{\alpha-1}}{\sqrt{A_{\alpha}^2v_0^2 + 4 K_\alpha s^\alpha}}
                 \exp\left(\frac{A_{\alpha}v_0 x}{2 K_\alpha} 
                          - |x| \frac{\sqrt{A_{\alpha}^2v_0^2+4 K_\alpha s^\alpha}}{2 K_\alpha}\right).
\end{equation}
In this case, a solution in $(x,t)$ space can be given as a superposition of the $\alpha = 1$
Gaussian solution with a L\'{e}vy kernel~\cite{Barkai2001, MeKl00}. However, for numerical
analysis we apply a direct numerical Laplace inversion of Eq.~\ref{eq:FFPE-C-xs}.

\subsubsection*{Type A and B FFPE:}
Analogously to Eq.~\ref{eq:FFPE-C-xs} the solutions of type A and type B FFPEs can 
be calculated in $(x,s)$ space with $A_{\alpha}v_0 \rightarrow v_0 s^{\alpha-1}$ to
\begin{equation}
\label{eq:FFPE-A+B-xs}
\tilde{f}_{A,B}(x,s) = \frac{s^{\alpha-1}}{\sqrt{v_0^2 s^{2\alpha-2}+ 4 K_\alpha s^\alpha}}
                 \exp\left(\frac{v_0 s^{\alpha-1} x}{2 K_\alpha} 
                          - |x| \frac{\sqrt{v_0^2 s^{2\alpha-2}+4 K_\alpha s^\alpha}}{2 K_\alpha}\right).
\end{equation}
As the FFPEs of type A and type B are Galilean invariant in the
  long time and large space limit, the solution for $v_0 = 0$ allows
the exact calculation of the PDFs with drift $v_0$ in this
  limit \cite{MKS98,MeKl00}, which becomes approximate otherwise
  \cite{CaBa15}. The solution to Eq.~\ref{eq:FDE2nodrift} is
  well-known~\cite{MeKl00} and is given using a Fox $H$-function
  (see~\ref{appendix:B} for definitions). Thus, applying Galilean
  transformation and replacing $x$ with $x-v_0 t$ gives solutions of
  type A and type B FFPEs in $(x,t)$ space as
\begin{equation}
\label{eq:fox-typeA+B}
f_{A,B}(x,t) = \frac{1}{\sqrt{4 K_\alpha t^\alpha}} \;
H^{1 0}_{1 1}
\left[ \frac{|x-v_0 t|}{\sqrt{K_\alpha t^\alpha}} \left| 
\begin{matrix}\displaystyle
(1-\alpha/2, \alpha/2)\\
(0,1) 
\end{matrix}
\right.
\right] .
\end{equation}
These approximate solutions in terms of shifted Fox functions are
the basis for our further analysis of type A and B FFPEs.

\subsection{Numerical analysis of time-fractional Fokker-Planck equations}
Numerical methods are required to study the analytical results given 
in form of Fox $H$-functions of type A and type B FFPE and in Laplace space
for type C FFPE.

\subsubsection*{Type A and type B FFPE:} The series expansion 
of the solution $f_{A,B}(x,t)$ of Eq.~\ref{eq:fox-typeA+B} 
as given by Eq.~\ref{eq:Fox1011-series-expansion}
is used for numerical evaluations,
\begin{equation}
f_{A,B}(x,t) = \frac{1}{\sqrt{4 K_\alpha t^\alpha}}
\sum_{j=0}^\infty \frac{(-1)^j}{j! \Gamma(1-\alpha(j+1)/2)} 
\left( \frac{(x-v_0 t)^2}{K_\alpha t^\alpha} \right)^{j/2}
\end{equation}
with $1<\alpha<2$ for type A FFPE and $0<\alpha<1$ for type B FFPE.
The series is evaluated with multiple-precision arithmetic.

\subsubsection*{Type C FFPE:} Direct numerical Laplace inversion is applied to  
Eq.~\ref{eq:FFPE-C-xs} to obtain the probability density function $f_C(x,t)$.  
Here we use a multiple-precision algorithm for the Laplace inversion 
based on Talbot's method~\cite{Talbot1979, Abate2004}.

\subsubsection*{Typical behavior in space and time:}

\begin{figure}[ht]
 \centering
 \includegraphics[width=0.9\linewidth]{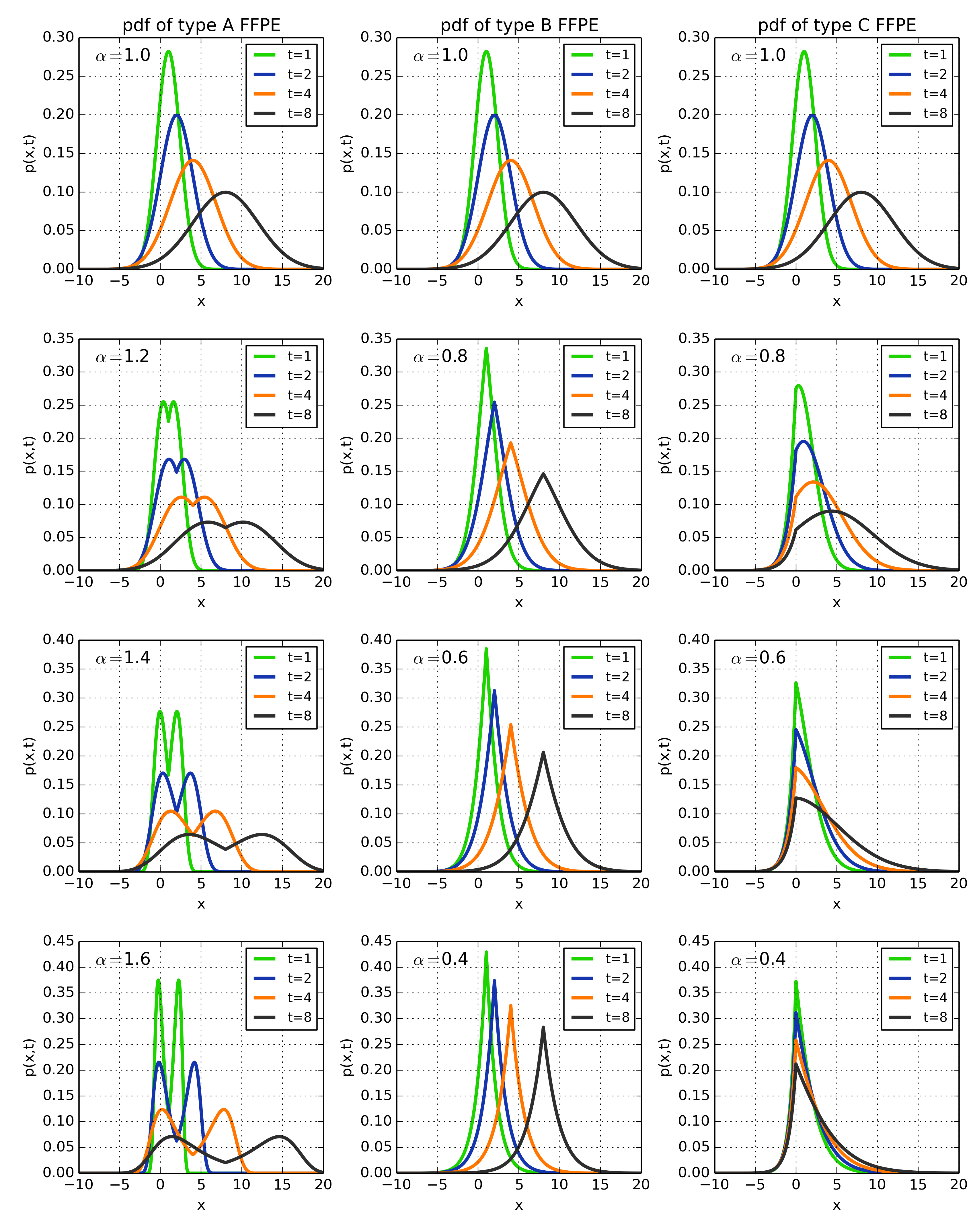} 
 \caption{\label{fig:FDEpdfs}Time development of PDFs for type A FFPE
 (left column), type B FFPE (middle column) and type C
 FFPE (right column) for different values of $\alpha$ (rows) and time
 points $t=1,2,4,8$. Parameters were selected as $K_\alpha = 1$, $v_0
 = 1$ and $A_{\alpha}v_0=1$. Whereas superdiffusive type A FFPE (left column)
 and subdiffusive type B FFPE (middle column) show a drift and spreading of
 the PDFs with typical non-Gaussian structures for $\alpha \neq 1$,
 type C FFPE (right column) displays a spreading of the PDFs together with
 stickiness to the origin.}
\end{figure}

Fig.~\ref{fig:FDEpdfs} shows the time development of the solutions
$f(x,t)$ of the three FFPE types for different times $t = 1, 2, 4,
8$. Parameters were selected as $A_{\alpha}v_0 = 1$ and $K_\alpha =
1$, the anomaly index $\alpha$ was chosen from $\alpha \in (0.4,0.6
... 1.6)$. The first row shows the Gaussian limit $\alpha \rightarrow
1$ for all three types. In this normal diffusive case the PDF is
spreading with $\sqrt{2 K_1 t}$ and its center is moving according to
$v_0 t$. The PDFs of type A (left column) and type B FFPE (middle
column) preserve this constant drift for $\alpha \neq 1$. However, the
shapes of the PDFs of both models immediately change profoundly
showing characteristically different types of non-Gaussian behavior:
For type A the PDFs spread superdiffusively with the variance of
Eq.~\ref{eq:FFPE-B-moment2c} by exhibiting a double-peaked structure
with a dip in the middle. Qualitatively, the highly characteristic
double-peak structure is explained in \cite{Bal07}: The propagator of
type A decays asymptotically {\em faster} than the Gaussian,
cf.~Eq.~\ref{eq:fox-asymptotics}. However, since two maxima move away
from the origin in the opposite directions, superdiffusion is possible
in spite of the thin tail of the propagator; see also
Eq.~\ref{eq:FDE+RLderiv} \cite{Metzler2000e}. Note that there
are cusp singularities in all three models for $\alpha \neq 1$, in
contrast to the smooth behavior of the Gaussian PDF shown in the top
row. In the Galilean invariant cases A and B the propagators are
symmetric with respect to their cusps, which are translated with
velocity $v_0 = 1$, as it should be. For the Galilean non-invariant
model C the propagator is asymmetric with respect to its cusp, which
stays fixed at the origin \cite{MeKl00}.

\section{Work fluctuation relations for fractional Fokker-Planck equations}

\subsection{Definition of fluctuation relations}
Using the results of the previous section, we now study the probability 
distribution $p(W,t)$ of the mechanical work $W = -F_0 \; x$ generated by the 
constant external field $F_0$. For a constant field the probability 
distribution of work $p(W,t)$ is related to the probability distribution $f(x,t)$ 
of positions by the simple scaling transformation
\begin{equation}
\label{eq:pdf-work-definition}
p(W,t) = \frac{1}{F_0} f\left(\frac{W}{F_0},t\right).
\end{equation}
It is the main aim of this work to study the TFR
of the work PDFs defined by the logarithmic fluctuation ratio
\begin{equation}
\label{eq:gen-R-def}
{\mathsf {R}}(W,t) := \log \frac{p(W,t)}{p(-W,t)}
\end{equation}
for the three types of FFPEs. All three FFPE types reduce to a normal 
Gaussian process with drift for $\alpha \rightarrow 1$. For a Gaussian PDF
the ratio ${\mathsf {R}}$ is trivially given by the ratio of the first and second central moment, i.e. 
$2 \langle W \rangle / \langle \delta W^2 \rangle$ \cite{ChKl12}. Thus one obtains a 
normal or conventional fluctuation relation for $\alpha = 1$,
\begin{equation}
{\mathsf {R}}(W,t)\left|_{\alpha=1}\right. = \frac{v_0}{F_0 K_1} \; W = W/(k_B T).
\end{equation}
with a linear increase in $W$ that is independent of time as it has been found for a large class of systems \cite{EvSe02,BLR05,HaSch07,JaPiRB11,Seif12,KJJ13}.
The last expression has been obtained by using the Einstein relation
$K_\alpha = k_B T/(m \gamma_\alpha)$ with temperature $T$, Boltzmann constant $k_B$
and the definition $v_0 = F_0/(m \gamma_\alpha)$. The general case for $\alpha \neq 1$
is studied in the next subsection.

\subsection{Fluctuation relations for fractional Fokker-Planck equations}

\subsubsection*{Type C FFPE:}
For this type the fluctuation ratio can be studied 
analytically~\cite{ChKl09}. With Eq.~\ref{eq:FFPE-C-xs}
${\mathsf {R}}(W,t)$  is given in Laplace space by
\begin{equation}
\label{eq:LT-TFR-FDE1}
\frac{\tilde{p}_C(W,s)}{\tilde{p}_C(-W,s)} =  
\frac{\tilde{f}_C(W/F_0,s)}{\tilde{f}_C(-W/F_0,s)} = \exp\left(\frac{A_{\alpha}v_0}{F_0 K_\alpha} W \right).
\end{equation}
As the right side is independent of the Laplace variable $s$, the Laplace inverse of
the PDFs can be calculated directly after multiplication with $\tilde{p}_C(-W,s)$.
Thus, despite the complicated form of the PDFs a linear normal TFR is obtained
for type C FFPE:
\begin{equation}
\label{eq:TPR-FFPE-C}
\log \frac{{p}_C(W,t)}{{p}_C(-W,t)} =  \frac{A_{\alpha}v_0}{F_0 K_\alpha} W .
\end{equation}
This result based on the Laplace transformed ratio of
$\tilde{p}_C(W,s)$ seems to be surprising with respect to the complex
form of the PDF in Laplace space and the asymmetric sticking behavior
at the origin of the PDFs as illustrated in the right column of
Fig.~\ref{fig:FDEpdfs}. The right side of Fig.~\ref{fig:FDE-FT} shows
the numerical calculation of the fluctuation ratio which is linear and
constant for all times in agreement with the given analytical result.

We remark that a normal TFR for type C can also be obtained with the
use of the subordination principle: Indeed, it is known that the
fractional kinetic equation C can be derived from the coupled Langevin
equations for the motion of a particle \cite{Fog94,BaFr05,ChKl09}
\begin{equation}
\label{eq:langevin_sub}
\frac{dx(u)}{du} = \frac{F_0}{m \gamma} + \xi(u)\:,\:\frac{dt(u)}{du}=\tau(u)\quad ,
\end{equation}
where the random walk $x(t)$ is parameterized by the random variable
$u$. The random process $\xi(u)$ is a white Gaussian noise,
$\langle\xi(u)\rangle=0\:,\:\langle\xi(u)\xi(u')\rangle=2k_bT\delta(u-u')/(m\gamma)$, and
$\tau(u)$ is a white stable L\'evy noise, which takes positive values
only and obeys a totally skewed $\alpha$-stable L\'evy distribution
with $0<\alpha<1$. The PDF $f(x,t)$ of the process $x(t)$ is then given
by
\begin{equation}
f(x,t)=\int_0^{\infty} du f_1(x,u)h(u,t)\quad ,
\end{equation}
where $f_1(x,u)$ is a shifted Gaussian PDF with drift, and $h(u,t)$
is the inverse one-sided L\'evy stable density \cite{Barkai2001}. It
is then easy to show that the linear normal TFR
Eq.~\ref{eq:TPR-FFPE-C} holds due to Gaussianity of $f_1$. Moreover, it
becomes clear that the normal TFR also holds for a more general form
of the PDFs $h(u,t)$, that is, for a more general class of the
positively valued stochastic processes $\tau(u)$.

\subsubsection*{Type A and B FFPEs:}
\begin{figure}[ht]
 \centering
 \includegraphics[width=0.9\linewidth]{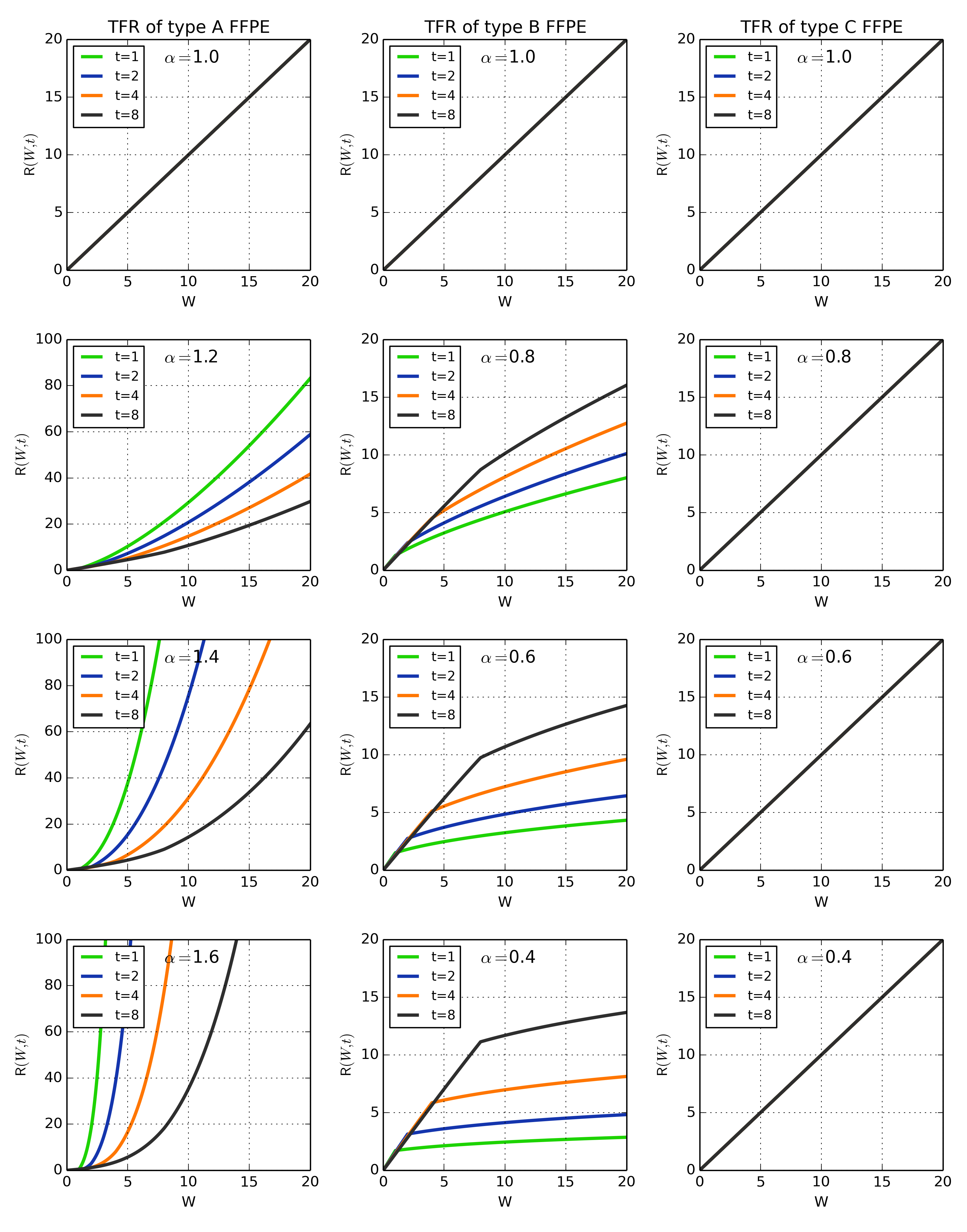} 
\caption{\label{fig:FDE-FT}Time dependence of the fluctuation ratio
for type A FFPE (left column), type B FFPE (middle column)
and type C FFPE (right column) for different values of $\alpha$
(rows) and times $t = 1, 2, 4, 8$. Parameters were selected as
$K_\alpha = 1$, $v_0 = 1$ and $A_{\alpha}v_0=1$. Whereas $\alpha = 1$
and all cases of $\alpha$ for type C FFPE show a normal fluctuation
ratio with time-independent slope (in all of these cases the linear
$t=8$ curve hides the previous times $t=1,2,4$) all other sub-plots
show a more complex time- and work-dependent fluctuation ratio:
Anomalous non-Markovian dynamics and/or non-Gaussian behavior cause a
complicated time-dependence and non-linear behavior of the work
fluctuation ratio.}
\end{figure}

For these two types the fluctuation ratio in Laplace space is more
complicated than for type C FFPE in Eq.~\ref{eq:LT-TFR-FDE1}. It is obtained 
with Eq.~\ref{eq:FFPE-A+B-xs} as
\begin{equation}
\label{eq:LT-TFR-FDE2}
\frac{\tilde{p}_{A,B}(W,s)}{\tilde{p}_{A,B}(-W,s)} =  
\frac{\tilde{f}_{A,B}(W/F_0,s)}{\tilde{f}_{A,B}(-W/F_0,s)} = \exp\left(\frac{v_0}{F_0 K_\alpha} s^{\alpha-1} W \right).
\end{equation}
In contrast to Eq.~\ref{eq:LT-TFR-FDE1}, here the right hand side
depends on the Laplace variable $s$. Consequently, one may expect an
anomalous ratio ${\mathsf {R}}$ which is confirmed numerically in the
overview of Fig.~\ref{fig:FDE-FT}. The fluctuation ratios of type A
(left column) and type B FFPEs (middle column) show a nonlinear
increase as functions of $W$. For type B FFPEs there is a clear
transition at the current maximum of the PDFs at $W_{max} = v_0 F_0 t$
which is equal to $t$ with $v_0 = 1$ and $F_0 = 1$ in
Fig.~\ref{fig:FDE-FT}.  For $W > W_{max}$ the fluctuation ratio
increases with time. In contrast, the TFR of type A FFPE increases
faster in $W$ than for type B. At the scale of this overview plot
there is no transition point visible as for type B FFPE. However, the
qualitative time-dependence of the fluctuation ratio for type A FFPE
is the opposite to type B FFPE: The ratio increases faster for smaller
times. To gain further insight into this behavior, some asymptotic
expansions of the TFR for type A and type B FFPEs are performed in the
next section.

\subsection{Asymptotic expansions of the fluctuation ratio for type A
and B FFPE} In this subsection we analyze the asymptotic behavior of
the work fluctuation ratio for type A and B FFPE. Differences between
type A and type B simply correspond to the value of $\alpha$ which is
$1<\alpha<2$ for the superdiffusive FFPE of type A and $0 <\alpha < 1$
for the subdiffusive type B FFPE. Type C is not considered anymore, as
the analytical calculation of Eq.~\ref{eq:TPR-FFPE-C} and the
numerical analysis in Fig.~\ref{fig:FDE-FT} have delivered a normal
fluctuation relation with a time-independent linear increase in the
work $W$.

\subsubsection*{Small $W$ expansion:} 
\begin{figure}[]
 \centering
 \includegraphics[width=0.9\linewidth]{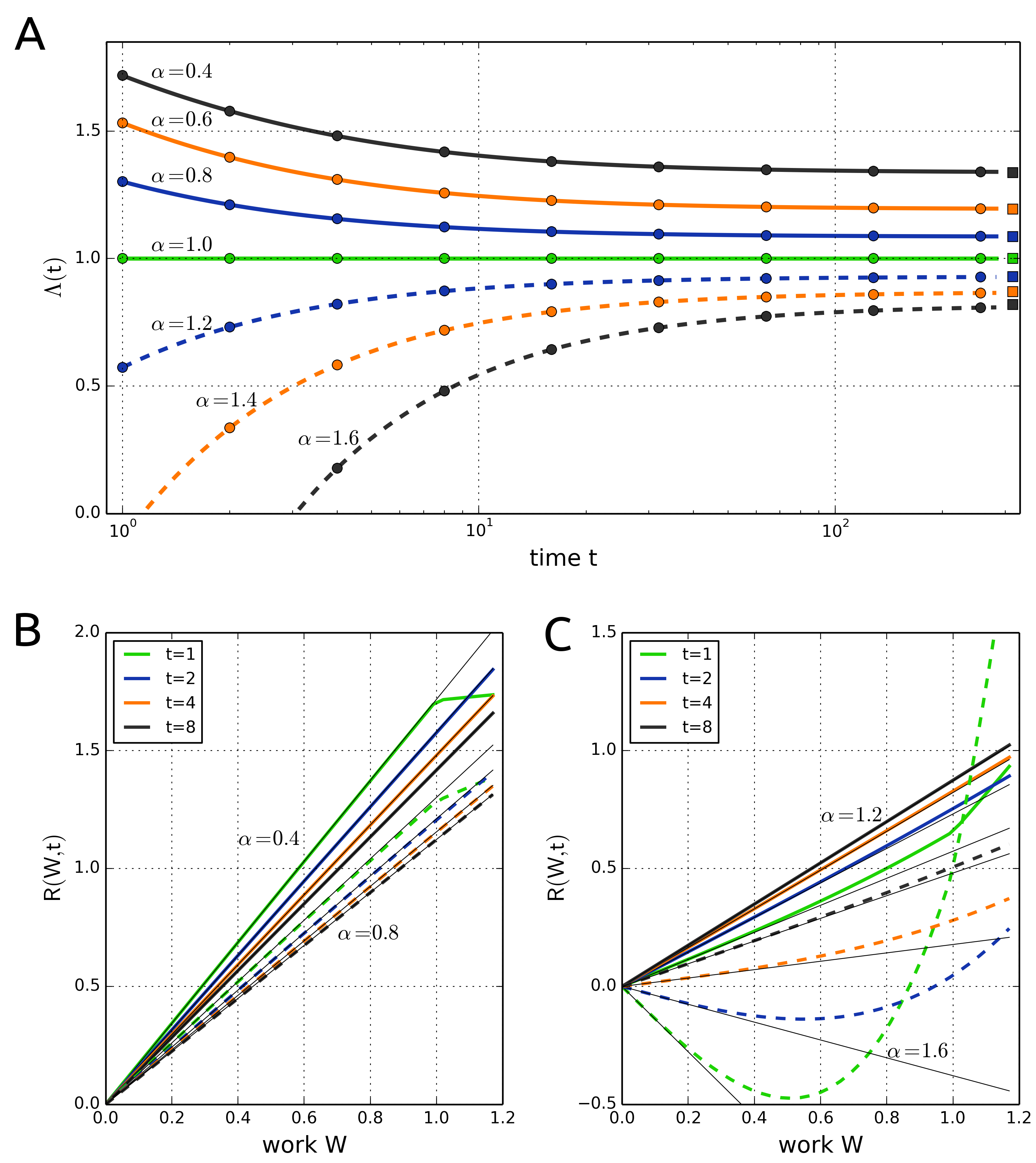} 
 \caption{\label{fig:FDE-FTdecay}{\bf A.} Time dependent decay of the initial fluctuation
 	      ratio $\Lambda(t)$ defined by Eq.~\ref{eq:TFR-FFPE-A+B-smallW} for small work $W$ and different values of $\alpha$ corresponding to
 	      type A FFPE ($1<\alpha<2$) and type B FFPE ($0<\alpha<1$) with parameters 
 		  $K_\alpha = 1$, $F_0 = 1$, $v_0 = 1$ and $A_{\alpha}v_0 = 1$. Circles show the direct calculation
 		  for small $W$ from the ratio of PDFs as defined in Eq.~\ref{eq:gen-R-def}
 		  whereas lines result from the computation of the first term of the small 
 		  $W$ expansion of Eq.~\ref{eq:logexpansion} and Eq.~\ref{eq:TFR-FFPE-A+B-smallW}.
 		  Both calculations agree and $\Lambda(t)$ converges towards 
 		  the long time limit given by Eq.~\ref{eq:Lambda-limit} as indicated
 		  by the squares. Whereas $\Lambda(t)$ is time-independent for $\alpha = 1$, 
 		  it decrease or increases as a function of time for the subdiffusive (type B FFPE)
 		  and superdiffusive case (type A FFPE), respectively. {\bf B.} The fluctuation
 		  ratio of work is shown for the subdiffusive case as a function of work and 
 		  different time points as indicated. The slope decreases for increasing time.
 		  Thin black lines indicate the small work limit of Eq.~\ref{eq:TFR-FFPE-A+B-smallW}. 
	          The obvious kink at $W=1$ for $t=1$ is due to the peak of the corresponding PDF in Fig.~\ref{fig:FDEpdfs}.
 		  {\bf C.} The superdiffusive case shows a more complicated behavior: The small
 		  work slope increases with time. In addition, it also changes from negative
 		  to positive for small time in the $\alpha = 1.6$ case.
 		  }
\end{figure}

First, the behavior of the TFR for the work 
PDFs of the FFPEs is studied for small $W$ as a function of time. The logarithmic 
ratio of a continuously differentiable function $p(z)$ can be expanded as 
Taylor series for positive $z$ as
\begin{equation}
\label{eq:logexpansion}
\log \frac{p(z)}{p(-z)} = \frac{2}{p(z=0)} \left. \frac{d p(z)}{dz} \right|_{z=0} z + \mathcal{O}(z^2).
\end{equation}
Inserting the approximate work PDF $p(W,t)$ from Eq.~\ref{eq:fox-typeA+B} together with the 
transformation of Eq.~\ref{eq:pdf-work-definition} into Eq.~\ref{eq:logexpansion} 
requires the calculation of the derivative of the Fox $H$-function. Using 
Eq.~\ref{eq:fox-derivative} with $r=1$, $h = 1$, $c=-1/\sqrt{K_\alpha t^{\alpha}}/F_0$, 
and $d = v_0 t/\sqrt{K_\alpha t^{\alpha}}$ allows us to calculate the linear term
in the Taylor expansion of Eq.~\ref{eq:logexpansion}. With the assumption $W/F_0 < v_0 t$ and 
after some simplifications using the definition of the Fox $H$-function
by the Mellin-Barnes integral in Eq.~\ref{eq:fox-def1} one obtains the
fluctuation ratio for small $W$ as a quotient of two Fox $H$-functions:
\begin{equation}
\label{eq:TFR-FFPE-A+B-smallW}
\left. \mathsf{R}(W,t)\right|_{W \rightarrow 0} = \left(\frac{2}{v_0 t}\right) \;
\frac{
	\displaystyle
	H^{1 0}_{1 1}
	\left[ \frac{v_0 t}{\sqrt{K_\alpha t^{\alpha}}} \left| 
	\begin{matrix}\displaystyle
	(1-\alpha/2,\alpha/2) \\
	(1,1) 
	\end{matrix}
	\right.
	\right]
}
{
	\displaystyle
	H^{1 0}_{1 1}
	\left[ \frac{v_0 t}{\sqrt{K_\alpha t^{\alpha}}} \left| 
	\begin{matrix}\displaystyle
	(1-\alpha/2,\alpha/2) \\
	(0,1) 
	\end{matrix}
	\right.
	\right]
}  \; \frac{W}{F_0}  = \Lambda(t) \; W.
\end{equation}
The prefactor $\Lambda(t)$ summarizes the time-dependence of the fluctuation ratio.
Its numerical evaluation based on the Taylor series of Eq.~\ref{eq:Fox1011-series-expansion}
is shown in Fig.~\ref{fig:FDE-FTdecay}A. In the superdiffusive case $1<\alpha<2$ (type A FFPE)
the prefactor $\Lambda(t)$ increases as a function of time, whereas in the 
subdiffusive case it decreases with time. The argument of the Fox $H$-functions 
$z=v_0 t /\sqrt{K_\alpha t^\alpha}$ in Eq.~\ref{eq:TFR-FFPE-A+B-smallW} scales 
$\sim t^{1-\alpha/2}$ with $1-\alpha/2 > 0$ for $0<\alpha<2$. Thus the asymptotic 
expansion of these Fox $H$-functions can be used for $t\rightarrow \infty$. In the long time 
limit the scaling function $\Lambda(t)$ converges towards the following non-zero 
constant value:
\begin{equation}
\label{eq:Lambda-limit}
\lim_{t \rightarrow \infty} \Lambda(t) = 
2 \left(\frac{2}{\alpha}\right)^{\alpha/(\alpha-2)} 
\left(\frac{v_0}{K_\alpha^{1/\alpha}}\right)^{\alpha/(2-\alpha)} \; \frac{W}{F_0}
\rightarrow \frac{v_0}{K_1 F_0} W
\;\;\;\text{for}\;\;\;\alpha=1.
\end{equation}
The corresponding values are shown as squares in Fig.~\ref{fig:FDE-FTdecay}A indicating
the predicted asymptotic behavior. Fig.~\ref{fig:FDE-FTdecay}B shows the spatial behavior
of the work fluctuation ratio for two subdiffusive examples $\alpha = 0.4$ and $0.8$ at
different instants of time $t$ (compare to small $W$ values in the overview given in 
Fig.~\ref{fig:FDE-FT}). The slope of the ratio decreases with increasing time and 
agrees well with the small $W$ expansion given in Eq.~\ref{eq:TFR-FFPE-A+B-smallW}. The 
superdiffusive case in Fig.~\ref{fig:FDE-FTdecay}C shows a reverse behavior as the small
$W$ ratio increases with time. As indicated in Fig.~\ref{fig:FDE-FTdecay}A it can also
be negative as show in Fig.~\ref{fig:FDE-FTdecay}C for $\alpha = 1.6$ and $t=1,2$. In 
the superdiffusive case, the small $W$ expansion has a smaller region of agreement with
the exact ratio. The more complex behavior is technically due to the two separating peaks
of the PDF as illustrated in Fig.~\ref{fig:FDEpdfs}.

\subsubsection*{Large $W$ expansion:}
\begin{figure}[]
 \centering
 \includegraphics[width=0.9\linewidth]{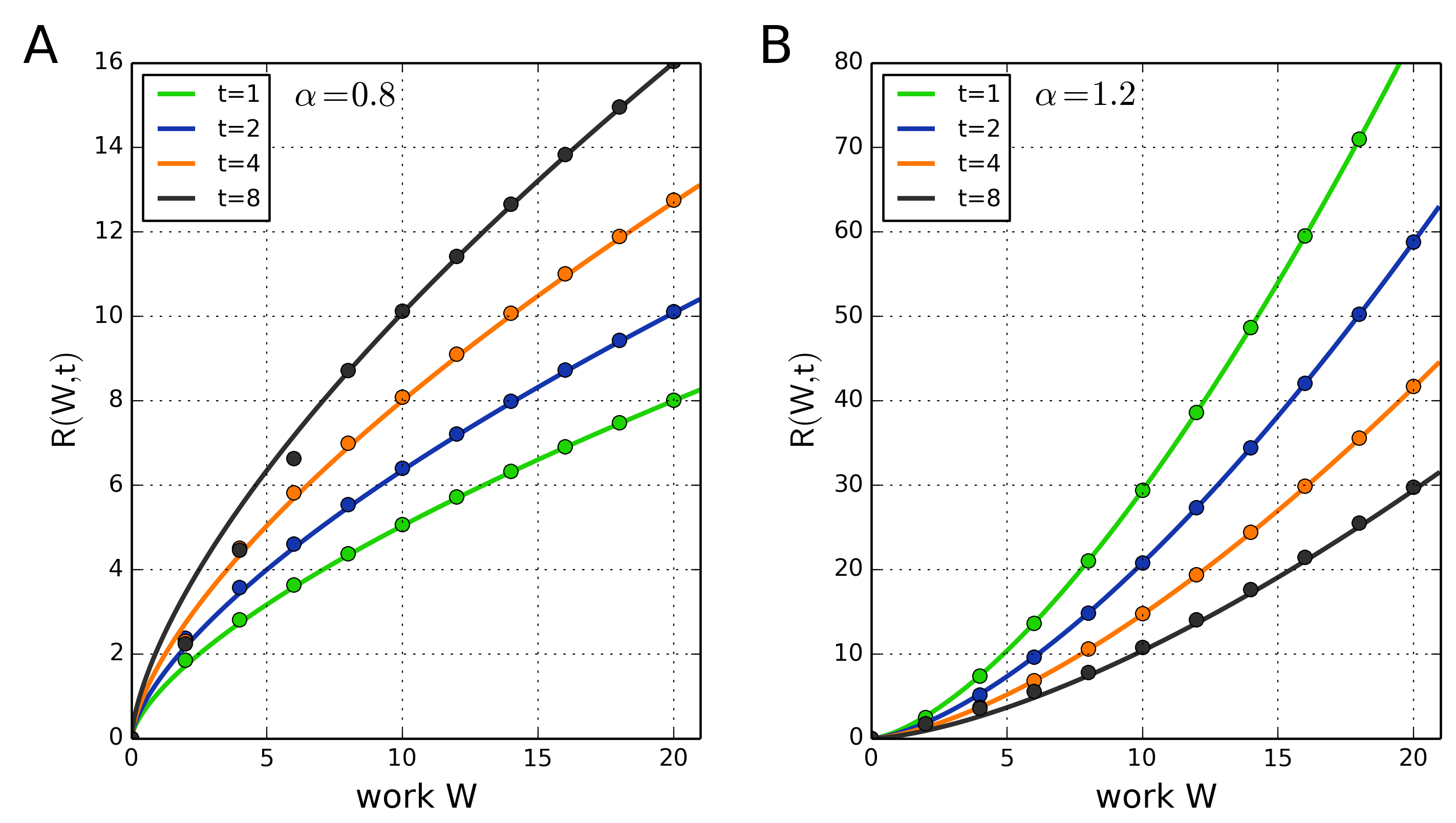} 
 \caption{\label{fig:FDE-FT-W-asymp} Large $W$ asymptotic of the work fluctuation 
 		  ratio of type A and type B FFPEs. Continuous lines show the asymptotic
 		  large $W$ result given by Eq.~\ref{eq:FT-W-asymp}. Circles indicate the
 		  exact result from the direct computation of the work fluctuation ratio.
 		  {\bf A.} Subdiffusive case for $\alpha = 0.8$ corresponding to type B
 		  FFPE. {\bf B.} Superdiffusive case for $\alpha = 1.2$ as example for
 		  type A FFPE. 
         }
\end{figure}

Finally, the behavior of the work fluctuation ratio is studied for large values
of the work $W$. The overview given in Fig.~\ref{fig:FDE-FT} shows a different non-linear
behavior for the subdiffusive and superdiffusive case. Assuming $W/F_0 > v_0 t$ and
large arguments of the Fox $H$-function for type A and type B FFPE in Eq.~\ref{eq:fox-typeA+B}
allows us to use the asymptotic expansion of the corresponding Fox $H$-function in 
Eq.~\ref{eq:fox-asymptotics}. For large $W$ one obtains the following relation:
\begin{equation}
\label{eq:FT-W-asymp}
\left. \mathsf{R}(W,t)\right|_{W \rightarrow \infty} =
\frac{2 v_0 t}{F_0} \; \left(\frac{\alpha}{2}\right)^{\alpha/(2-\alpha)} 
\left( \frac{1}{\sqrt{K_\alpha t^\alpha}} \right)^{2/(2-\alpha)} \; W^{\alpha/(2-\alpha)} .
\end{equation}
Thus the work fluctuation ratio scales as a power law with an exponent $\alpha/(2-\alpha)$.
This exponent is between $0$ and $1$ for the subdiffusive type B FFPE. For superdiffusive
type A FFPE it is larger than $1$. This asymptotic power law behavior is shown in 
Fig.~\ref{fig:FDE-FT-W-asymp} for two examples. Continuous lines represent the result 
of Eq.~\ref{eq:FT-W-asymp} and agree for larger $W$ values with the exact results denoted
by circles. Eq.~\ref{eq:FT-W-asymp} additionally contains a time-dependent scaling
factor that is proportional $t^{(2\alpha-2)/(\alpha-2)}$. This factor is positive for the 
subdiffusive type B FFPE and negative for type A FFPE.

\section{Summary and outlook}

In this work we studied three different types of FFPEs generating
anomalous diffusion: a superdiffusive one (type A), a subdiffusive one
(type B), and another one that exhibits a transition from sub- to
superdiffusion under parameter variation (type C). Type A and type B
break FDR1 while type C preserves it. Type A can be derived, under
certain assumptions, from an overdamped Langevin equation with power
law correlations of the velocity fluctuations, types B and C have been
derived before in the literature from CTRW theory. Type C can also be
obtained via subordination. We then calculated position PDFs for all
models analytically and studied the shapes of all PDFs numerically
under variation of the anomaly index as they evolve in time. Finally
we checked the work TFR for all three models. Especially, we studied
the time dependence of the ratio of the work fluctuations both for
small and for large work by analytical asymptotic expansions in
comparison to numerical evaluations.

We find that our type C model with FDR1 exhibits a conventional work
TFR for all times, meaning the fluctuation ratio is constant in time
and linear in the work. For a correlated Gaussian stochastic process
it was shown that FDR1 implies the existence of a conventional TFR
\cite{ChKl12}. Our work generalises this result to an example of
non-Gaussian PDFs generated by FFPE dynamics. It is interesting that
the conventional TFR is still obeyed, despite the highly non-trivial
dynamics exhibited by both the position PDFs and the corresponding
moments. The existence of the conventional TFR for this case is
connected to the fact that only the equation for Type C describes a
subordinated process, namely the one subordinated to Brownian motion
with drift under random time transformation. An important open
question is to which extent Fig.~\ref{fig:FDEpdfs} in \cite{ChKl12}
summarising the interplay between FDR1, FDR2 and TFRs for correlated
Gaussian stochastic processes in terms of necessary and sufficient
conditions can be generalised to non-Gaussian processes. For our other
two models type A and type B the position PDFs show also very subtle
and non-trivial non-Gaussian shapes. However, in contrast to type C
they are characterised by a highly non-trivial fluctuation ratio: For
type A the latter decreases with time, for type B it
increases. Similar results have been obtained for the work TFR of
strongly correlated Gaussian stochastic processes without FDR1
\cite{ChKl09,ChKl12}. On top of this, for both types of FFPEs the
fluctuation ratio yields different long time limits depending on
whether the work is small or large: For small work the fluctuation
ratio converges to linearity in the work with constant prefactors,
which reminds of the conventional TFR; however, here the slopes depend
on the anomaly index of the dynamics. For large work the fluctuation
ratio remains nonlinear in the work, with convex and concave shapes
for type A and type B, respectively.

Our work was motivated by experiments on cell migration \cite{DKPS08},
where data were successfully fitted by solutions of a fractional
Klein-Kramers equation \cite{BaSi00}. Several generalisations of such
a Klein-Kramers equation have been proposed to describe processes
under external fields \cite{BaSi00,MeSo02,FJBE06}, which in turn yield
FFPEs for the position only, similar to the ones studied in our paper,
as special cases \cite{Lutz01,SchnWy89,MeKl00}. We thus believe that
our present work might have important applications to understand cell
migration in nonequilibrium situations such as under chemical
gradients; see \cite{KCD13} for first results. More generally, our
theory might have applications to understand glassy nonequilibrium
dynamics: In computer simulations of a number of glassy systems
violations of conventional TFRs have been observed featuring
fluctuation ratios that are nonlinear in the work with time-dependent
prefactors \cite{Sell09,CPR13}. 

Apart from such experimental applications, our first approach for
deriving a FFPE pioneered by Balescu \cite{Balescu1997,Bal07} deserves
to be studied in more detail. For example, it would be interesting to
derive a superdiffusive FFPE from it that preserves FDR1, and to check
again the TFR. On a broader scale it would be important to generalise
our approach by considering more general observables, ideally
dissipation functions \cite{EvSe02} or related functionals defined
within stochastic thermodynamics \cite{Seif12}. More general force
fields than simply constant forces \cite{ChKl09} and other types of
FRs could be tested as well. Such theoretical studies may pave the way
to identify different classes of anomalous FRs characterized by
specific functional forms, generalized FDRs associated with them, and
to explore the physical significance of these results. Last not
  least the quality of the Galilean invariant approximate solution
  Eq.~\ref{eq:fox-typeA+B} \cite{MKS98,MeKl00} of the FFPEs
  \ref{eq:FFPE-A},\ref{eq:FFPE-B} needs to be investigated in detail.

\section*{Acknowledgment}

We thank A.Cairoli for very helpful discussions.

\appendix
\section{\label{appendix:A}Pseudo-Liouville approach}
Following the so-called pseudo-Liouville hybrid approach of Balescu~\cite{Balescu1997,Bal07}
allows us to relate the dynamics of a particle defined by a Langevin equation to the
corresponding PDF of the stochastic process. We start from the Langevin equation for 
the position $x(t)$ of a particle
\begin{equation}
\label{eq:simple-langevin}
\frac{dx(t)}{dt} = v_0 + v(t)\:,
\end{equation}
where $v(t)$ is a correlated stochastic process with zero mean $\langle v(t) \rangle = 0$
and a given correlation function $\langle v(t) v(t') \rangle = \mathsf{T}(t-t')$, where the
average is performed over the stochastic process $v(t)$. $v_0$ 
denotes a constant external force. The stochastic function $F(x,t)$
\begin{equation}
\label{eq:exact-density}
F(x,t) = \delta(x-x(t))
\end{equation}
represents the exact density of the process. Derivation of Eq.~\ref{eq:exact-density}
with respect to time and the usage of the Langevin equation Eq.~\ref{eq:simple-langevin}
delivers the continuity equation for the exact density $F(x,t)$:
\begin{equation}
\label{eq:F-continuity}
\frac{\partial F(x,t)}{\partial t} = 
-\frac{\partial}{\partial x} \delta(x-x(t)) \; \frac{dx(t)}{dt}
\longrightarrow
\frac{\partial F(x,t)}{\partial t} + [v_0 + v(t)] \frac{\partial F(x,t)}{\partial x} = 0.
\end{equation}
Now, the exact density $F(x,t)$ is decomposed into an averaged part $f(x,t)$ and fluctuations
$\delta f(x,t)$
\begin{equation}
\label{eq:F-decomposition}
F(x,t) = f(x,t) + \delta f(x,t) 
\;\;\; \mbox{with} \;\;\;
f(x,t) = \langle F(x,t) \rangle .
\end{equation}
It is the further aim of this appendix to calculate the PDF $f(x,t)$ for the stochastic 
process defined by the Langevin equation Eq.~\ref{eq:simple-langevin} for given correlations
of $v(t)$. Averaging of the exact density in Eq.~\ref{eq:F-continuity} leads to
\begin{equation}
\label{eq:f-df-eq}
\left(\frac{\partial}{\partial t} + v_0 \frac{\partial}{\partial x}\right) f(x,t)
= -\frac{\partial}{\partial x} \big\langle v(t) \delta f(x,t) \big\rangle .
\end{equation}
Subtraction of Eq.~\ref{eq:f-df-eq} from Eq.~\ref{eq:F-continuity} results in
\begin{equation}
\label{eq:df-f-eq}
\left(\frac{\partial}{\partial t} + [v_0 + v(t)] \frac{\partial}{\partial x}\right) \delta f(x,t)
= -\frac{\partial}{\partial x} 
\bigg( v(t)\;f(x,t) - \big\langle v(t) \; \delta f(x,t) \big\rangle \bigg) .
\end{equation}
Eq.~\ref{eq:df-f-eq} can be solved with the method of characteristics 
\begin{equation}
\label{eq:df-solution}
\delta f(x,t) = -\frac{\partial}{\partial x}
\int\limits_0^t dt' 
\bigg( 
	v(t') \; f(x-\Delta(t,t'),t')  
	- \big\langle v(t') \; \delta f(x-\Delta(t,t'),t') \big\rangle 
\bigg)
\end{equation}
with the definition $\Delta(t,t') = v_0 (t-t') - \int\limits_{t'}^t dt_1 \; v(t_1)$. Inserting
Eq.~\ref{eq:df-solution} into Eq.~\ref{eq:f-df-eq} delivers the final equation for the 
PDF $f(x,t)$:
\begin{equation}
\label{eq:f-solution}
\left(\frac{\partial}{\partial t} + v_0 \frac{\partial}{\partial x}\right) f(x,t)
=
\frac{\partial^2}{\partial x^2}
\int\limits_0^t dt' 
\big\langle	v(t) v(t') \; f(x-\Delta(t,t'),t') \big\rangle . 
\end{equation}
This is an exact relation for $f(x,t)$ that is generally non-local in space and non-local
in time, i.e. non-Markovian. Applications and approximations of this relation are studied 
in section~\ref{sec:TFDE.Balescu}.

\section{\label{appendix:B}Definiton and properties of Fox $H$-functions}
The Fox $H$-function is defined as inverse Mellin transform of the function 
$\chi(s)$~\cite{MeKl00,Mainardi2005}
\begin{equation}
\label{eq:fox-def1}
H_{p,q}^{m,n}(z) 
= H_{p,q}^{m,n}\left[ z \bigg|\bigg. 
                              \begin{matrix}(a_j,A_j)_{j=1,p}\\(b_j,B_j)_{j=1,q}\end{matrix}      
                              \right] 
= \frac{1}{2\pi i} \int_{\cal L} \chi(s) z^s ds
\end{equation}
over a suitable path $\cal L$, with
\begin{equation}
\label{eq:fox-def2}
\chi(s) = \frac{\prod\limits_{j=1}^m \Gamma(b_j-B_j s) \; \prod\limits_{j=1}^n \Gamma(1-a_j+A_j s)}
               {\prod\limits_{j=m+1}^q \Gamma(1-b_j+B_j s) \; \prod\limits_{j=n+1}^p \Gamma(a_j-A_j s)},
\end{equation}
$0 \le n \le p$, $1 \le m \le q$, $(a_j, b_j) \in \mathbb{C}.$, and $(A_j, B_j) \in \mathbb{R}^+$. Empty
products in Eq.~\ref{eq:fox-def2} are taken as one.

A series expansion allows the numerical calculation of Fox $H$-functions. The following form for
a special Fox $H$-function is used:
\begin{equation}
\label{eq:Fox1011-series-expansion}
H^{1 0}_{1 1}
\left[ z \left| 
\begin{matrix}\displaystyle
(a_1,A_1) \\
(b_1,B_1) 
\end{matrix}
\right.
\right]
=
\sum\limits_{k=0}^{\infty} \; 
\frac{(-1)^k}{\Gamma\left(\displaystyle a_1-A_1\frac{b_1+k}{B_1} \right)}\;
\frac{\displaystyle z^{\frac{b_1+k}{B_1}}}{k!\;B_1} .
\end{equation}
Summation in this work is performed numerically with multiple-precision arithmetic. 

The derivation of the Fox $H$-function is required to calculate the fluctuation
ratio for the FFPEs of type A and B. This can be performed
using the following relation~\cite{Mathai2010}:
\begin{eqnarray}
\label{eq:fox-derivative}
\frac{d^r}{d x^r} 
H^{m,n}_{p,q} &&
\left[ (c x + d)^h \left| 
\begin{matrix} \displaystyle
(a_j, A_j)_{1,p} \\
(b_j, B_j)_{1,q}
\end{matrix}
\right.
\right] = 
\nonumber \\
&&= 
\left( \frac{c}{c x + d} \right)^r \;
H^{m,n+1}_{p+1,q+1}
\left[ (c x + d)^h \left| 
\begin{matrix}\displaystyle
(0,h), (a_j, A_j)_{1,p} \\
(b_j, B_j)_{1,q}, (r,h)
\end{matrix}
\right.
\right] .
\end{eqnarray}
For large arguments the Fox $H$-functions of type $H^{q,0}_{p,q}(z)$ decay as stretched exponential
functions. The asymptotics of the PDF in Eq.~\ref{eq:fox-typeA+B} is given for large $z$ by~\cite{Braaksma1962,Mathai2010}
\begin{equation}
\label{eq:fox-asymptotics}
f_{A,B}(z,t) \sim \frac{z^{-\frac{1-\alpha}{2-\alpha}}}{\sqrt{4 \pi (2-\alpha) K_\alpha t^\alpha}}
\left(\frac{2}{\alpha}\right)^{\frac{1-\alpha}{2-\alpha}} \;
\exp\bigg\lbrace
	-\left(\frac{2-\alpha}{2}\right) \; \left(\frac{2}{\alpha}\right)^\frac{\alpha}{\alpha-2} \; z^{\frac{2}{2-\alpha}}      
	\bigg\rbrace 
\end{equation}
for $z \rightarrow \infty$ and $z = |x-v_0 t|/\sqrt{K_\alpha t^\alpha}$.

\section*{References}

\end{document}